\documentclass[12pt]{article}

\def\be{\begin{equation}\label}
\def\ee{\end{equation}}
\def\bea{\begin{eqnarray}\label}
\def\eea{\end{eqnarray}}
\def\a{\alpha}
\def\b{\beta}

\begin{document}

\author{Fabrizio Canfora,  Hans-J\"urgen Schmidt}
\title{Vacuum solutions which cannot be written in diagonal 
form\footnote{Address: Institut f\"ur Mathematik, Universit{\"a}t
Potsdam,  Am Neuen Palais 10, D-14469 Potsdam, Germany. \ 
E-mail:  fabrcanf@tiscalinet.it 
\ and \ hjschmi@rz.uni-potsdam.de 
\quad http://www.physik.fu-berlin.de/\~{}hjschmi}}
\date{}
\maketitle

\begin{abstract}
The vacuum solution 
\[
ds^2 = dx^2 + x^2 \, dy^2 + 2 \, dz \, dt + \ln x \, dt^2 
\]
of the Einstein gravitational field equation follows from the general ansatz 
\[
ds^2 = dx^2 + g_{\alpha \beta} (x) \, dx^\a dx^\b 
\]
but fails to follow from it if the symmetric matrix $g_{\alpha \beta} (x) $
is assumed to be in diagonal form.
\end{abstract}

\smallskip

\noindent KEY: Vacuum solution, Einstein field equation, symmetries,
diagonalization

\medskip 

\section{Introduction}

The folklore reading ``Every symmetric matrix can be brought into diagonal
form by a suitable rotation'' is strictly valid in the positive definite
case only. In the Lorentz signature case, however, one needs additional
assumptions to get this result. It is generally believed that these
assumptions do not represent a real restriction, and this is justified,
e.g., for the energy-momentum tensor: all physically sensible form of matter
can be written with an energy-momentum tensor in diagonal form.

\bigskip

It is the purpose of the present paper to show, that, nevertheless,
important examples exist where this folklore-statement leads to incorrect
results. Of course, some examples of this kind already exist. Most notably,
the Kerr metric, and all its generalizations, cannot be brought into
diagonal form
in holonomic coordinates due to the fact that its timelike Killing field
fails to be hypersurface-orthogonal. However, there is a widespread feeling
that, in the case of the metric depending only on one coordinate, this
diagonalization can always be achieved. Here we give a class of solutions
where this rough diagonalization is not possible. This is essentially
important if one wishes to find {\it all} solutions of a prescribed symmetry
type. In this respect, this is a continuation of \cite{x1} and \cite{x4}.

\bigskip

The paper is organized as follows: Section 2 presents the deduction of the
Bianchi type I vacuum solutions of Einstein's gravitational field equations
in such a detailed form that it becomes clear, why the Kasner solution \cite
{x2}, here cited from \cite{x3},\footnote{%
According to \cite{x3}, the Kasner solution published in 1921 was already
known to Levi-Civita in 1917, cf. also \cite{x6}.} is really possible in
diagonal form.\footnote{%
To the question when symmetric matrices {\it cannot} be diagonalized see 
\cite{x5}.} Section 3 gives the analogous calculation as section 2, but now
with a changed signature. In the final section 4 we discuss the  case of
a metric with only one off-diagonal term. We will show that the vacuum
Einstein equations for this kind of metric, that is a system of four
nonlinear differential equations with four unknown functions, can always be
reduced to a system two equations with two unknown functions. Moreover, in
 some cases we will be able to further reduce the system to only one equation
in one unknown function. We will also show one explicit solution of the
system, that is a vacuum space-time which cannot be diagonalized.

\section{The Kasner solution}

The general metric for a Bianchi type I model reads 
\begin{equation}  \label{y1}
ds^2 = - dt^2 + g_{\alpha \beta} (t) \, dx^\a dx^\b
\end{equation}
where $g_{\alpha \beta} (x) \, dx^\a dx^\b $ is the positive definite
spatial metric. We want to find out all vacuum solutions of the Einstein
field equation of the form of metric~(\ref{y1}). The result reads, cf.
section 11.3. of \cite{x3}: 
\begin{equation}  \label{y2}
ds^2 = - dt^2 + t^{2p} dx^2 + t^{2q} dy^2 + t^{2r} dz^2
\end{equation}
with 
\begin{equation}  \label{y3}
p+q+r = p^2+q^2+r^2 \in \{ 0, \, 1 \} \ .
\end{equation}
Proof:\footnote{%
It is a fully standard proof, but here we need the details to see the
distinction to the other cases to be discussed below.} This is a well-posed
Cauchy problem. We take $[ t=0]$ as initial space-like hypersurface. It is a
3-flat space. Therefore, we can take without loss of generality 
\begin{equation}  \label{y4}
g_{\alpha \beta} (0) = \delta_{\alpha \beta} = \left( 
\begin{array}{c}
1 \quad 0 \quad 0 \\ 
0 \quad 1 \quad 0 \\ 
0 \quad 0 \quad 1
\end{array}
\right) \, ;
\end{equation}
otherwise a coordinate transformation involving only the 3 spatial
coordinates $x^1 = x$, $x^2 = y$, $x^3 =z$ suffices to reach that.

\bigskip

The second fundamental tensor at $[ t=0]$ is $k_{\alpha \beta}(0)$, where 
\begin{equation}  \label{y5}
k_{\alpha \beta} = \frac{d}{dt} \, g_{\alpha \beta}
\end{equation}
represents a symmetric matrix. The remaining freedom of spatial coordinate
transformations keeping valid the equation (\ref{y4}) is just the orthogonal
group $O(3)$, see Appendix A for the proof that the 3 parameters of $O(3)$
suffice\footnote{%
In fact, we have proven even more: already the connected component of the
unity element of $O(3)$, namely the $SO(3)$, is enough to get that result,
but we do not need this additional property here.} to reach 
\begin{equation}  \label{y6}
k_{12}(0) = k_{13} (0) = k_{23}(0) =0 \, .
\end{equation}
Therefore, due to the compactness of the rotation group we can assume
without loss of generality that the second fundamental tensor at $[ t=0]$
has diagonal form.

\bigskip

The vacuum Einstein field equations are equivalent to $R_{ij}=0$, where $%
R_{ij}$ represents the Ricci tensor, and the index of the coordinates $x^{i}$
covers the values $i=0,1,2,3$, and $t=x^{0}$. Then it turns out that the 3
equations 
\begin{equation}
R_{12}=R_{13}=R_{23}=0  \label{y7}
\end{equation}
suffice to maintain the diagonal form of the metric $g_{\alpha \beta }(t)$
for all times.

\bigskip

Up to now we have shown that all vacuum metrics of the form (\ref{y1}) can
be written as 
\begin{equation}  \label{y8}
ds^2 = - dt^2 + e^{2\alpha(t)} dx^2 +e^{2\beta(t)} dy^2 + e^{2\gamma (t)}
dz^2 \, .
\end{equation}
Inserting this metric into $R_{ij} =0$ it turns out that up to trivial
rescalings, the one-parameter set of solutions defined by eqs. (\ref{y2}, 
\ref{y3}) cover the set of all solutions.

\bigskip

In the final step one observes, that $-dt^2 + dx^2$ and $-dt^2 + t^2 dx^2$
are both locally flat, and therefore, one may omit the case $p=q=r=0$ from
eq. (\ref{y3}) without loosing any solutions.

\bigskip

Result: Every cosmological Bianchi type I solution of the Einstein vacuum
field equations, i.e., every solution of the form (\ref{y1}) can be written
as (\ref{y2}) with\footnote{%
A geometric parametrization of this set is given in Appendix B.} 
\begin{equation}
p+q+r=p^{2}+q^{2}+r^{2}=1\ .  \label{y9}
\end{equation}

\section{The signature changed Kasner solution}

In this section we want to deduce the consequences of another signature in
the metric. First of all, one would be tempted to go just the same way as
before: Looking at eqs. (\ref{y2}, \ref{y3}), one can perform an imaginary
rotation of $x$ and $y$. After rewriting $ds^{2}$ as $-ds^{2}$ and renaming
the coordinates one gets: 
\begin{equation}
ds^{2}=dx^{2}+x^{2r}dy^{2}+x^{2q}dz^{2}-x^{2p}dt^{2}  \label{y10}
\end{equation}
It holds: If eq. (\ref{y9}) is fulfilled, then metric (\ref{y10}) represents
a vacuum solution of the Einstein field equations. Contrary to the positive
definite case, a permutation between $p$, $q$, and $r$ is no more generally
possible. It remains only the permutation between $q$ and $r$. So, the set
of solutions in metric (\ref{y10}) can be parametrized by eqs. (\ref{b6}, 
\ref{b5}, \ref{b4}) with $0\leq \phi \leq \pi $.

\bigskip

We will now carefully look for the question whether these solutions
represent all solutions of the form of metric (\ref{y1}).

\subsection{The diagonal ansatz}

The diagonal ansatz analogous to eq. (\ref{y8}) reads 
\begin{equation}  \label{y11}
ds^2 = dx^2 + e^{2\gamma(x)} dy^2 +e^{2\beta(x)} dz^2 - e^{2\alpha (x)} dt^2
\, .
\end{equation}
Inserting this metric into the equation $R_{ij} =0$, one gets as expected,
again just the known solutions (\ref{y10}) with (\ref{y9}).

\subsection{The non-diagonal ansatz}

We are now only interested to show that truly non-diagonal metrics really
exist, and we do not intend to exhaust all of them in the present paper.
Therefore, we restrict to those metrics, where only one off-diagonal element
of $g_{\alpha \beta} (x) $ is different from zero.

\bigskip

An off-diagonal component between two space-like directions can be made
vanish by the procedure shown in Appendix A. So, this essential off-diagonal
component must exist between one space-like and one time-like direction.
This leads us to the following ansatz for the metric: 
\begin{equation}
ds^{2}=dx^{2}+A(x)dy^{2}+g_{mn}(x)dx^{m}dx^{n}  \label{y12}
\end{equation}
where $A(x)>0$ and $g_{mn}(x)$ is negative definite. We count the
coordinates $x^{3}=z$ and $x^{4}=t$, so the indices $m,\,n$ run from 3 to 4.
We write the 2-dimensional metric 
\begin{equation}
g_{mn}(x)=\left( 
\begin{array}{c}
B(x)\quad P(x) \\ 
P(x)\quad -C(x)
\end{array}
\right)   \label{y13}
\end{equation}
and use the abbreviation: $-\det g_{mn}=\Gamma =P^{2}+BC$. The conditions
for the negative definiteness are: 
\[
A>0  \  \,  \  \Gamma >0 \, .
\]
The inverse reads 
\begin{equation}
g^{mn}(x)=-\frac{1}{\Gamma }\left( 
\begin{array}{c}
-C(x)\quad \ P(x) \\ 
\ P(x)\quad -B(x)
\end{array}
\right) \,.  \label{y14}
\end{equation}

\section{The Einstein equations}

\bigskip

Since we are dealing with the vacuum case, the Einstein equations reduce to 
\[
R_{i j }=0 {\bf ,} 
\]
where $R_{i j }$ is the Ricci tensor of the metric (\ref{y12}). It is well
known that,
due to the Bianchi identities, not all the Einstein equations are
independent. In this case it is convenient to take, as our basic equations: 
\[
R_{yy}=0 {\bf , } \ R_{zz}=0 {\bf , } \ R_{tt}=0 {\bf , } \ R_{tz}=0 {\bf . } 
\]
It is a straightforward calculation to compute the Ricci components. The
explicit expressions read: 
\begin{eqnarray}
R_{yy} =0\Rightarrow -2\Gamma A\stackrel{..}{A}+\Gamma \left( \stackrel{.}{%
A}\right) ^{2}-A\stackrel{.}{A}\stackrel{.}{\Gamma }=0 {\bf ,}  \label{e1} \\
R_{zz} =0\Rightarrow 2\Gamma \stackrel{..}{B}+2B\left( \stackrel{.}{P}%
\right) ^{2}-\stackrel{.}{B}\left( 2P\stackrel{.}{P}+C\stackrel{.}{B}-B%
\stackrel{.}{C}-\Gamma \frac{\stackrel{.}{A}}{A}\right) =0 {\bf ,}
\label{e2} \\
R_{tt} =0\Rightarrow 2\Gamma \stackrel{..}{C}+2C\left( \stackrel{.}{P}%
\right) ^{2}-\stackrel{.}{C}\left( 2P\stackrel{.}{P}+B\stackrel{.}{C}-C%
\stackrel{.}{B}-\Gamma \frac{\stackrel{.}{A}}{A}\right) =0 {\bf ,}
\label{e3} \\
R_{tz} =0\Rightarrow 2\Gamma \stackrel{..}{P}+2P\stackrel{.}{B}\stackrel{.%
}{C}-\stackrel{.}{P}\left( B\stackrel{.}{C}+C\stackrel{.}{B}-\Gamma \frac{%
\stackrel{.}{A}}{A}\right) =0 {\bf ,}  \label{e4}
\end{eqnarray}
where $\stackrel{.}{f}=\frac{df}{dx}$. First of all, let us notice that eq. (%
\ref{e1}) can be explicitly solved for $A$: 
\begin{eqnarray}
2A^{\frac{1}{2}} &=&\kappa ^{\frac{1}{2}}\int_{0}^{x}\frac{dx^{\prime }}{%
\sqrt{\Gamma }}+I_{1}\Rightarrow  \label{A} \\
&\Rightarrow &\frac{\kappa }{\stackrel{.}{A}}=\Gamma \frac{\stackrel{.}{A}%
}{A}  \label{sostA}
\end{eqnarray}
where $I_{1}$ and $\kappa $ are integration constants. Now, since $A$ is
expressed in terms of $\Gamma $, it is clear that, thanks to the identity (%
\ref{sostA}), we are left with a system of three equations in the three
unknown functions $B$, $C$ and $P$. This system looks highly nontrivial due
to the nonlinearities. Nevertheless, it is possible to further reduce it by
rewriting the equations (\ref{e2}), (\ref{e3})  and (\ref{e4})  in a more
symmetric
way. It is important to stress here that, thanks to (\ref{e2})  and
(\ref{e3})%
, $P^{2}$ is a symmetric function in the exchange of $B$ and $C$. Hence, $P$
can be either symmetric or antisymmetric in the exchange of $B$ and $C$. In
the following $B$, $C$ and $P$ will be supposed to be different from zero. In
fact, if $P$ is zero one obtains a Kasner-like solution, while if $B$, or $C$%
, is zero then it is always possible to make a coordinate transformation in
such a way that $B\neq 0$, the same holds for $C$. Now, it is easy to show
that the
system of equations (\ref{e2}), (\ref{e3}) and (\ref{e4}) is equivalent to the
following system: 
\begin{equation}
2\Gamma \frac{\stackrel{..}{Y_{i}}}{Y_{i}}+2\Sigma -\frac{\stackrel{.}{Y_{i}}%
}{Y_{i}}\left( \stackrel{.}{\Gamma }-\frac{\kappa }{\stackrel{.}{A}}%
\right) =0 {\bf ,}  \label{f1}
\end{equation}
where $Y_{1}=B$, $Y_{2}=C$, $Y_{3}=P$ and $\Sigma =\left( \stackrel{.}{P}%
\right) ^{2}+\stackrel{.}{B}\stackrel{.}{C}$. It could look that in these
equations we cannot put $\stackrel{.}{A}=0$. However, by remembering the
identity (\ref{sostA}), it is obvious that one obtains $\stackrel{.}{A}=0$
by taking in these equations $\kappa =0$. Thus, we rewrote  the system
of the equations (\ref{e2}), (\ref{e3})  and (\ref{e4})  in a manifestly
symmetric
form: all the $Y_{i}$'s obey the same equation. By introducing the
variables $%
\eta _{ij}=\stackrel{.}{Y}_{i}Y_{j}-\stackrel{.}{Y}_{j}Y_{i}$, i.e. the
Wronskian of $Y_{i}$ and $Y_{j}$, and supposing $\stackrel{.}{\eta }%
_{ij}\neq 0$ $\forall i,j$ (otherwise, if, for some $i$ and $j$, $\stackrel{.%
}{\eta }_{ij}=0\Rightarrow Y_{i}\sim Y_{j}$ and the system is immediately
reduced) we arrive at the following system of equations: 
\begin{equation}
2\Gamma \stackrel{.}{\eta }_{ij}-\left( \stackrel{.}{\Gamma }-\frac{%
\kappa }{\stackrel{.}{A}}\right) \eta _{ij}=0 {\bf .}  \label{w1}
\end{equation}
From this system, it immediately follows that 
\[
\frac{\stackrel{.}{\eta }_{ij}}{\eta _{ij}}=\left( \log \eta _{ij}\right)
^{.}=\frac{\left( \stackrel{.}{\Gamma }-\frac{\kappa }{\stackrel{.}{A}}%
\right) }{2\Gamma } {\bf ,} 
\]
so, after a trivial integration, it comes out that the $\eta _{ij}$'s are all
proportional: 
\begin{equation}
\eta _{12}\sim \eta _{13}\sim \eta _{23} {\bf .}  \label{wronprop}
\end{equation}
As it is well known from the theory of the linear system of ordinary
differential equations, that (\ref{wronprop}) implies that one of the three
unknown functions is a linear combination of the other two. It is convenient
to choose $P$ as dependent function. In fact, since we know that $P^{2}$ is
symmetric in the exchange of $B$ and $C$, then the only possibilities for $P$
are: 
\[
P=\alpha \left( B\pm C\right)  {\bf ,} 
\]
where $\alpha $ is an arbitrary nonzero constant. In this way, we reduced
the initial system of four nonlinear equations in four unknown functions to
a system of two equations in the unknown functions $B$ and $C$. In general,
due to the term $\frac{\kappa }{\stackrel{.}{A}}$ that couples in a 
non--trivial way $B$ and $C$, it is not possible either to decouple the two
equations or to further reduce the system. However, if one takes 
\begin{equation}
P=\pm \frac{1}{2}\left( B-C\right)  {\bf ,}  \label{pspecial}
\end{equation}
then the system (\ref{f1}) can be reduced to one equation in one unknown
function. In fact, if eq. (\ref{pspecial}) holds, then 
\[
\Gamma =\frac{1}{4}\left( B+C\right) ^{2} {\bf , }\Sigma =\frac{1}{4}\left( 
\stackrel{.}{B}+\stackrel{.}{C}\right) ^{2} {\bf .} 
\]
Now, if we introduce 
\[
u=B+C {\bf , }  \    v=B-C {\bf ,} 
\]
then, from eq. (\ref{f1}), $u$ and $v$ satisfy the following two equations: 
\begin{eqnarray}
\frac{u^{2}}{2}\stackrel{..}{u}+\frac{\stackrel{.}{u}^{2}}{2}u+\stackrel{.}{u%
}\left( \frac{\kappa }{\stackrel{.}{A}}-\frac{u}{2}\stackrel{.}{u}\right)
&=&0 {\bf ,}  \label{t1} \\
\frac{u^{2}}{2}\stackrel{..}{v}+\frac{\stackrel{.}{u}^{2}}{2}v+\stackrel{.}{v%
}\left( \frac{\kappa }{\stackrel{.}{A}}-\frac{u}{2}\stackrel{.}{u}\right)
&=&0 {\bf  .}  \label{t2}
% \\
%A^{\frac{1}{2}} &=&\kappa ^{\frac{1}{2}}\int_{0}^{x}\frac{dx^{\prime }}{u}%
%+I_{1} {\bf .}
\end{eqnarray}
Since eq. (\ref{t2}) is a linear homogeneous ordinary differential
equation, $%
v $ can be expressed in a closed form in terms of $u$, so the system
(\ref{f1}) 
is reduced to the only eq. (\ref{t1}). Hence, once eq. (\ref{t1}) is solved
for $%
u $, the other metric coefficients immediately follow.

\subsection{An explicit example}

An interesting explicit example is the following: 
$$
ds^{2} =dx^{2}+x^{2}dy^{2}+\left( 1+\frac{1}{2}\ln \left| x\right| \right)
dz^{2}-\left( 1-\frac{1}{2}\ln \left| x\right| \right) dt^{2}-\ln \left|
x\right| dz\,dt\,  ,  
$$
\begin{equation}
\  {\rm where  \ }\frac{1}{e} < \sqrt{\left| x\right| }<e \,  .  \label{y17} 
\end{equation}
In this case we have: 
\[
B+C=2 {\bf , }\  B-C=\ln \left| x\right|  {\bf , } \  P=-\frac{1}{2}\left(
B-C\right)  {\bf .} 
\]
Then it is trivial to show that $u=B+C$ satisfies eq. (\ref{t1}), while
$v=B-C$
satisfies eq. (\ref{t2}) for $\kappa =4$ and $A$ follows from eq. (\ref{A}) 
by taking $I_{1}=0$. It is interesting, at this point, to make a comparison
with the Kasner case. In particular, one could ask: why the procedure to
diagonalize the metric in the Kasner case works and in this case does  not
work? In this case, the first part of the exercise is the same as the Kasner
one: Let us take the initial hypersurface $[x=0]$, and then without loss of
generality let 
\begin{equation}
a(0)=1\quad {\rm and}\quad g_{mn}(0)=\left( 
\begin{array}{c}
1\quad \ 0 \\ 
0\quad -1
\end{array}
\right) \,.  \label{y15}
\end{equation}
Then we define 
\begin{equation}
k_{mn}=\frac{d}{dx}g_{mn}  \label{y16}
\end{equation}
and try to diagonalize $k_{mn}(0)$. However, due to the non-compactness of
the Lorentz group, the arguments of Appendix A no more apply; moreover, see
Appendix C, one can really find examples of matrices $k_{mn}(0)$ which
cannot be brought into diagonal form. The calculation to be done is
straightforward and will be omitted here. Then, it is clear that the
differences between the two cases are group theoretical in nature.

\section*{Appendix A}

Let 
\begin{equation}  \label{a1}
{\bf k} = k_{\alpha \beta} = \left( 
\begin{array}{c}
a \quad d \quad e \\ 
d \quad b \quad f \\ 
e \quad f \quad c
\end{array}
\right)
\end{equation}
be any symmetric matrix, i.e. ${\bf k } = {\bf k }^{{\rm T}}$. We want to
show that in the positive definite case, this matrix can be diagonalized;
the 3-dimensional real orthogonal group is denoted by $O(3)$, and the
superscript T denotes the transposed matrix. Then ${\bf U} \in O(3)$ acts
continuously on {\bf k} to give 
\begin{equation}  \label{a2}
{\bf U}^{{\rm T}} \cdot {\bf k} \cdot {\bf U}\, .
\end{equation}
We have to show that one can always choose ${\bf U} \in O(3)$ such that the
matrix eq. (\ref{a2}) has diagonal form. To this end we define the quantity 
\begin{equation}  \label{a3}
J( {\bf k}) = d^2 + e^2 + f^2 \, .
\end{equation}
Due to the compactness of $O(3)$, the minimum of $J$ exists; we have to
prove that this minimum leads to $J=0$. Assumed, this is not the case.
Without loss of generality we may assume that this is due to $d \ne 0$, for
otherwise, a permutation of the coordinate axes would lead to this
inequality.

\bigskip

Let ${\bf A}_{\phi} \in O(3) $ be defined by 
\begin{equation}  \label{a4}
{\bf A}_{\phi} = \left( 
\begin{array}{c}
\ \cos \phi \quad \ \, \sin \phi \quad 0 \\ 
-\sin \phi \quad \cos \phi \quad 0 \\ 
\quad 0 \quad \quad \ 0 \ \quad \quad \ 1
\end{array}
\right) \, .
\end{equation}
It holds: The inverse matrix to ${\bf A}_{\phi}$ equals ${\bf A}_{- \phi}$
which is nothing but ${\bf A}_{\phi}^{{\rm T}}$. Analogously to expression (%
\ref{a2}), we define 
\begin{equation}  \label{a5}
{\bf k}_\phi = {\bf A}_{\phi}^{{\rm T}} \cdot {\bf k} \cdot {\bf A}_{\phi}\,
,
\end{equation}
and then we get with eqs. (\ref{a3}) and (\ref{a4}) up to linear order in
the Taylor expansion with respect to $\phi$ 
\begin{equation}  \label{a6}
J( {\bf k}_\phi ) = J( {\bf k} ) + 2 (a-b) \cdot d \cdot \phi \, .
\end{equation}
For $a \ne b$ we are already finished: A small change of $\phi$ will change
the value of $J$ linearly with $\phi$, so there cannot be a minimum at $\phi
=0$. For $a=b$, this linear expansion does not suffice to decide, but here,
the exact value is easy to evaluate, it reads: 
\begin{equation}  \label{a7}
J( {\bf k}_\phi ) = J( {\bf k} ) - 4 \cdot d^2 \cdot \sin^2 \phi \cdot
\cos^2 \phi \, .
\end{equation}
There is a local maximum at $\phi =0$,
 and therefore, this cannot be a minimum. Result: the
assumption $d \ne 0$ leads to a contradiction. 

\section*{Appendix B}

Equation (\ref{y9}), i.e., 
\begin{equation}  \label{b1}
p+q+r = p^2+q^2+r^2 = 1
\end{equation}
represents the intersection of the plane $p+q+r =1$ with the unit sphere in
the $p-q-r-$space. Thus, it must be a circle. We parametrize it by the
angular coordinate $\phi$. The following 3 points 
\begin{equation}  \label{b2}
P=(1,\, 0, \, 0) \, , \quad Q=(0, \, 1, \, 0) \, , \quad R = (0, \, 0, \, 1 )
\end{equation}
are obviously on this circle; in turn, this circle is uniquely determined by
them. The center $M$ of this circle is given by the arithmetic mean of $P$, $%
Q$ and $R$, i.e. 
\begin{equation}  \label{b3}
M= (\frac{1}{3}, \, \frac{1}{3}, \, \frac{1}{3}) \, .
\end{equation}
Its radius equals the distance from $M$ to $P$, i.e. $\sqrt{2/3}$. So, we
get the parametrization of eq. (\ref{b1}) as 
\begin{eqnarray}  \label{b6}
p = \frac{1}{3} + \frac{2}{3} \, \cos \phi \\
q = \frac{1}{3} \left( 1 - \cos \phi \right) + \sqrt{1/3} \, \sin \phi
\label{b5} \\ \label{b4}
r = \frac{1}{3} \left( 1 - \cos \phi \right) - \sqrt{1/3} \, \sin \phi 
\end{eqnarray}
Obviously, it suffices to restrict to the $\phi$-interval $0 \le \phi < 2
\pi $. However, a permutation between the 3 numbers $p$, $q$ and $r$ can be
compensated by a coordinate transformation (namely a related permutation of
the spatial axes of metric (\ref{y2})), therefore, to get a one-to-one
correspondence it proves useful to require additionally $r \le q \le p$.
Comparing eqs. (\ref{b5}) and (\ref{b6}) one can see that the inequality $r
\le q$ is fulfilled for $0 \le \phi \le \pi$ only. The other inequality, $q
\le p$ further reduces this interval via the identity 
\begin{equation}  \label{b7}
p-q = \cos \phi - (\sin \phi) /\sqrt 3
\end{equation}
to 
\begin{equation}  \label{b8}
0 \le \phi \le \pi/3\, .
\end{equation}
Clearly, as there are six possible permutations, the length of this interval
is $2\pi/6$. The boundary of this interval consists of two points. The point
related to $\phi =0$ is the already discussed flat space-time. The other
one, related to $\phi = \pi/3$, i.e. that point where $p=q=2/3$,
 $r=-1/3$  is the other
axially symmetric solution for Bianchi type I. Here one can see what one
also meets in other circumstances: The solutions with higher symmetry (here:
axial symmetry) are at the boundary of the space of solutions.

\section*{Appendix C}

Let 
\begin{equation}  \label{c1}
{\bf \eta} = \left( 
\begin{array}{c}
1 \quad \ 0 \\ 
0 \quad -1
\end{array}
\right) \, .
\end{equation}
The Lorentz group $O(1,1)$ is the group of all those transformations leaving
the matrix eq. (\ref{c1}) invariant. For a given symmetric matrix 
\begin{equation}  \label{c2}
{\bf k} = \left( 
\begin{array}{c}
A \quad D \\ 
D \quad B
\end{array}
\right)
\end{equation}
defined by 3 parameters, the one-parameter group $O(1,1)$ acts continuously,
so, from counting the degrees of freedom one could be tempted to assume,
that one can always choose an element of $O(1,1)$ such that $D$ becomes
zero. However, this is not the case. For our purposes it suffices to give an
example: For 
\begin{equation}  \label{c3}
A=B=D=1/2
\end{equation}
put into eq. (\ref{c2}), no diagonalization is possible.

\bigskip

Let us look from another side: The trace of {\bf k}, namely the expression%
\footnote{%
The minus sign in front of $B$ is due to the minus 1 in eq. (\ref{c1}).} $%
A-B $, and the determinant, namely $AB-D^2$, are invariants of it with
respect to $O(1,1)$-actions. Again, the counting is misleading: 3 free
parameters in (\ref{c2}), a one-parameter gauge group, so these two
invariants should suffice for an invariant characterization. But this is not
true: both $A=B=D=1/2$ and $A=B=D=0$ lead to a vanishing of both invariants,
whereas no element of $O(1,1)$ can be given that transforms the one into the
other. This is analogous to the discussion in \cite{x7}: two objects are
different, but no continuous invariant exists to distinguish between them.
Here it holds: every continuous invariant of {\bf k} can be written as a
function of trace and determinant only.

\section*{Appendix D}

Let a metric be given as 
\begin{equation}  \label{d1}
ds^2 = dx^2 + x^2 \, dy^2 + 2 \, dz \, dt + a( x ) \, dt^2
\end{equation}
where $a(x)$ is any free function. Due to the off-diagonal term $dz \, dt$, $%
a(x)$ may have zeroes without leading to a singularity there.

\bigskip

We denote $(x, \, y, \, z, \, t)$ by $x^i$, $i=1, \dots 4$. The only
component of the Ricci tensor $R_{ij}$ which does not vanish identically, is 
\begin{equation}  \label{d2}
R_{44} = - \frac{1}{2} \left( \frac{d^2 a}{dx^2} + \frac{1}{x} \cdot \frac{da%
}{dx} \right) \, .
\end{equation}

\bigskip

The only components of the Riemann tensor $R_{ijkl}$ which do not vanish
identically, are 
\begin{equation}  \label{d3}
R_{1414} = - \frac{1}{2} \cdot \frac{d^2 a}{dx^2} \, , \qquad R_{2424} = - 
\frac{x}{2} \cdot \frac{da}{dx} \, .
\end{equation}
This statement is meant, of course, only ``up the usual antisymmetries".

\bigskip

As a result of eq. (\ref{d3}) we find: Metric (\ref{d1}) is flat if and only
if the function $a$ is a constant.

\bigskip

To find out all non-flat vacuum solutions of the Einstein field equation of
the form (\ref{d1}), one has therefore to solve $R_{44}=0$ using eq. (\ref
{d2}) with a nonconstant $a(x)$. The result is, after a possible
redefinition of the coordinates $t$ and $z$, be expressible as 
\begin{equation}  \label{d4}
a(x) = c \pm \ln x
\end{equation}
where $c$ is a given constant of integration.

\bigskip

Let us calculate the curvature invariants of metric (\ref{d1}): Let $I$ be
any polynomial invariant like $R^{ij} \, R_{ij}$. Then $I$ depends on the
one coordinate $x$ only. To calculate one special value $I(x)$ we make the
following construction: We replace, for any positive real $\epsilon$, the
coordinate $t$ by $\epsilon \, t$ and the coordinate $z$ by $z/ \epsilon$.
This does not change the form of metric (\ref{d1}), only the function $a(x)$
is now replaced by $\epsilon^2 \cdot a(x)$.

In the limit $\epsilon \to 0$, we meet the flat spacetime having $I \equiv 0$%
. But $I(x)$ is a continuous function, and as invariant it does not change
with $\epsilon$, therefore: Every polynomial curvature invariant for metric (%
\ref{d1}) identically vanishes.

\section*{Note added}

The paper [9] and  ours are different in scope, but the discussed 
metrics have much overlap: [9] presents the most general 
 metric that depends on just one coordinate and cannot be diagonalized. 
The metric is a generalization of
the Levi-Civita, or Kasner metrics.  The authors of [9] also analyzed 
the global structure
of the spacetime described by this metric -- it has closed timelike curves,
without a Cauchy horizon, and the question of whether such metric can
represent the ``exterior" of some ``tube of matter", the answer is that in
general an energy condition is violated.

\bigskip

In [10], several results on diagonalization procedures can be found, too. 
 According to M. MacCallum, the solution in the
abstract of the present paper is just a special pp-wave and appears as eq.
(22.5) 
in the new edition of ref. [5] (with $a=1$ and  $ \rho =  \ln x$). It also 
 represents a special case of solutions already given in refs. [1] and [4]. 
 In refs. [11] and [12],  similar solutions have been discussed, too.

\section*{Acknowledgement}

The authors are grateful to Professor H. Kleinert for hospitality at the 
Free University Berlin where this work has been done. F. C. also
thanks Professor G. Vilasi for continuous encouragement.
We thank G. Clement, A. Feinstein, M. MacCallum and M. Tiglio
for valuable comments after acceptance of this paper; their
comments are incorporated in the ``Note added" above.

\bigskip
\noindent 
[9] Reinaldo J. Gleiser, Manuel H. Tiglio: Exotic spacetimes, 
superconducting strings with linear momentum, and (not quite) 
all that; gr-qc/0001087;  Phys. Rev. {\bf D 61} (2000) 104006.

\bigskip
\noindent 
[10] M. MacCallum: Hypersurface-orthogonal generators of an      
orthogonally transitive
transitive $G_2I$, topological  identifications, 
and axially and cylindrically symmetric
     spacetimes, Gen. Rel. Grav. {\bf 30} (1998) 131.

\bigskip
\noindent 
[11] C. McIntosh: Real Kasner and related complex 
windmill vacuum spacetime metrics,  Gen. Rel. Grav. {\bf  24}, 757 
(1992).

\bigskip
\noindent 
[12] G. Clement, I. Zouzou: Hollow cosmic string: the general-relativistic
hollow cylinder, Phys. Rev.  {\bf D 50} (1994) 7271, gr-qc/9405074.
\end{document}